\documentclass[10pt,letterpaper,twocolumn,aps,showpacs,preprintnumbers,amsmath,amssymb,nofootinbib,prl]{revtex4-2}
\usepackage[latin9]{inputenc}
\usepackage{color}
\usepackage{amsmath}
\usepackage{amssymb}
\usepackage{graphicx}

\usepackage[unicode=true,pdfusetitle,
 bookmarks=true,bookmarksnumbered=false,bookmarksopen=false,
 breaklinks=false,pdfborder={0 0 0},pdfborderstyle={},backref=false,colorlinks=true]
 {hyperref}
\hypersetup{
 pdfborderstyle={},colorlinks,linkcolor=red,citecolor=blue}

\makeatletter

\usepackage{setspace}
\usepackage{dcolumn}
\usepackage{mathrsfs}
\usepackage{bm}
\usepackage{siunitx}
\usepackage{chngcntr}
\usepackage{multirow}
\usepackage{tabularx}

\makeatother

\begin{document}
\title{Quantum Interference between Photons and Single Quanta of Stored Atomic Coherence}

\author{Xingchang Wang,$^{1,2}$ Jianmin Wang,$^{1,2}$ Zhiqiang Ren,$^{4}$ Rong Wen,$^{5}$ Chang-Ling Zou,$^{6}$ Georgios A. Siviloglou,$^{1,2,\ast}$ and J. F. Chen$^{1,3,\dagger}$}

\affiliation{$^{1}$Shenzhen Institute for Quantum Science and Engineering and Department of Physics, Southern University of Science and Technology, Shenzhen 518055, China\\
$^{2}$International Quantum Academy (SIQA), and Shenzhen Branch, Hefei National Laboratory, Futian District, Shenzhen 518055, China\\
$^{3}$Guangdong Provincial Key Laboratory of Quantum Science and Engineering, Southern University of Science and Technology, Shenzhen 518055, China\\
$^{4}$State Key Laboratory of Precision Spectroscopy, School of Physics and Electronic Science, East China Normal University, Shanghai 200241, China\\
$^{5}$Key Laboratory of Advanced Transducers and Intelligent Control System of Ministry of Education, College of Physics and Optoelectronics, Taiyuan University of Technology, Taiyuan 030024, Shanxi, China\\
$^{6}$CAS Key Laboratory of Quantum Information, University of Science and Technology of China, Hefei 230026, Anhui, China\\
$^{\ast}$siviloglouga@sustech.edu.cn
$^{\dagger}$chenjf@sustech.edu.cn}
\date{\today}

\begin{abstract}
Essential for building quantum networks over remote independent nodes, the indistinguishability of photons has been extensively studied by observing the coincidence dip in the Hong-Ou-Mandel interferometer. However, indistinguishability is not limited to the same type of bosons. For the first time, we hereby observe quantum interference between flying photons and a single quantum of stored atomic coherence (magnon) in an atom-light beam splitter interface. We demonstrate that the Hermiticity of this interface determines the type of quantum interference between photons and magnons. Consequently, not only the bunching behavior that characterizes bosons is observed, but counterintuitively, fermionlike antibunching as well. The hybrid nature of the demonstrated magnon-photon quantum interface can be applied to versatile quantum memory platforms, and can lead to fundamentally different photon distributions from those occurring in boson sampling.
\end{abstract}
\maketitle

\begin{figure*}
\includegraphics[width=2.0\columnwidth]{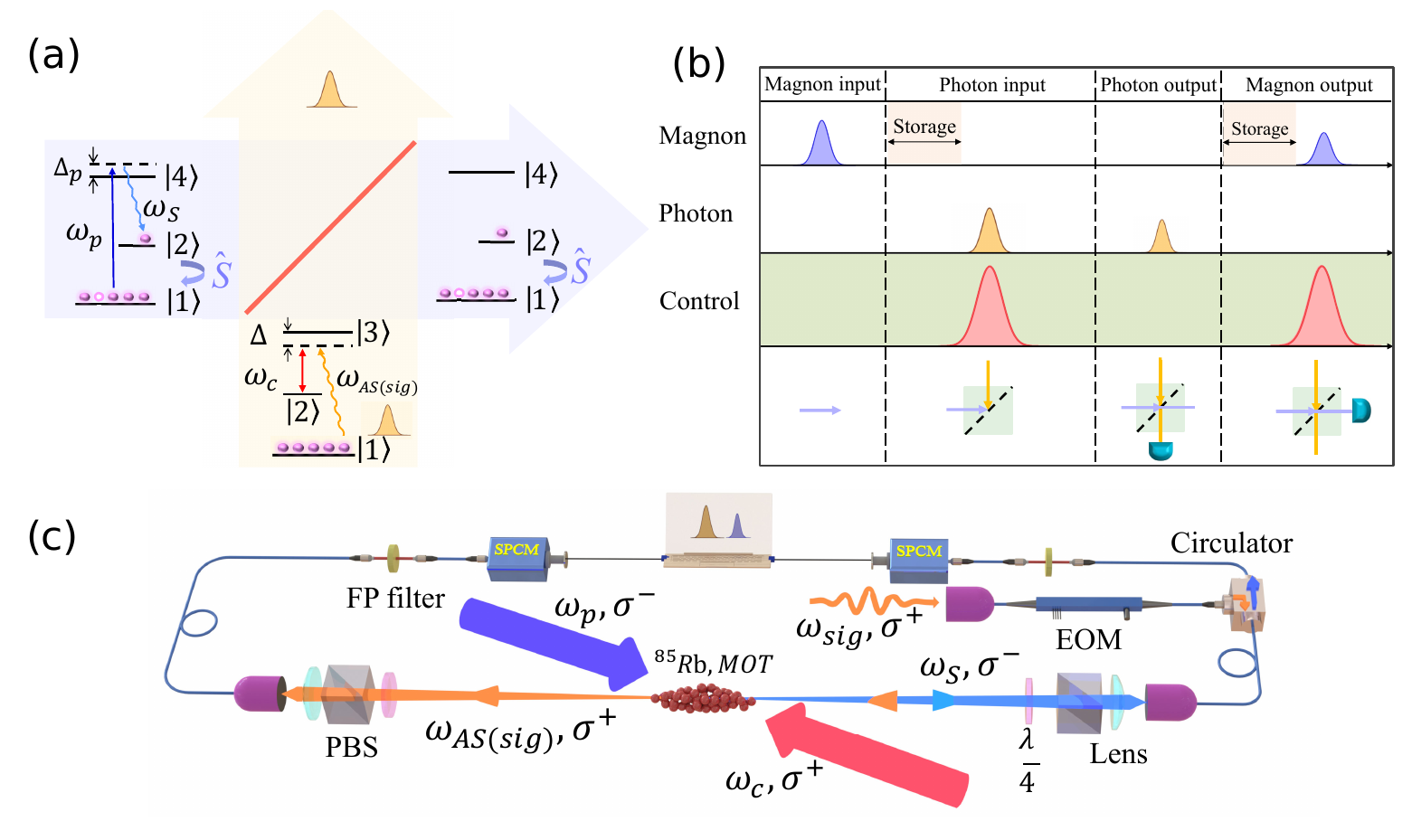} \caption{The experimental protocol for quantum interference in a non-Hermitian MPBS. (a) Energy scheme of the MPBS. A single magnon $\hat{S}$ between the two metastable states $|1\rangle:~5S_{1/2},F=2$ and $|2\rangle:~5S_{1/2},F=3$ is generated by a pump beam $\omega_p$ with detuning $\Delta_{p}$ from state $|4\rangle:~5P_{3/2},F=3$, while it is heralded by the spontaneous emission of a single Stokes photon $\omega_S$. A single-photon-level light pulse $\omega_{sig}$ enters the atomic medium and conversion between the photon and the magnon takes place in the presence of an EIT control beam $\omega_c$ between the states $|2\rangle$ and $|3\rangle:~5P_{1/2},F=3$ with two-photon resonance satisfied. The transmitted photons and the remaining magnons are detected as anti-Stokes photons $\omega_{AS}$. (b) Timeline of the quantum interference protocol. Starting from the left, the pump beam first creates the single magnons, photons enter the cold ensemble, and interference between magnons and photons takes place. The transmitted photons are detected as photon output, and the magnons are later read out by another control pulse. The storage steps before and after quantum interference are described together with Fig.~$\,$\ref{Fig4}. (c) Experimental setup with laser cooled $^{85}Rb$ atoms from a magneto-optical trap (MOT). The strong control and pump beams are circularly polarized. The input photons $\omega_{sig}$ are shaped by an electro-optic modulator (EOM), and the generated photons from $\omega_{S}$ and $\omega_{AS}$ are detected, after passing from polarization elements (PBS: polarization beam splitter) and Fabry-Perot (FP) filters, by single-photon counting modules (SPCM) in a time-resolved manner.}
\label{Fig1}
\end{figure*}

Quantum memories require faithful and efficient conversion between photonic and matter qubits~\cite{Lvovsky2009}. Typical schemes of optical information storage which harness atomic spin coherence have been demonstrated in various physical systems, such as ion-doped solid crystals~\cite{Raha2020,USTC2021NC} and neutral atom ensembles~\cite{Dudin2013,Korber2018}. In an ideal quantum memory, a single photon could be stored as a single quantum of atomic spin coherence, which is approximated as a quantized bosonic quasiparticle magnon, and then converted back to an indistinguishable photon in exactly the same mode as the original one. On-demand generation and retrieval of atomic spin coherence are realized with different coherent conversion mechanisms, including electromagnetically induced transparency (EIT)~\cite{Lukin2000, Hsiao2018PRL,yunfei2019}, off-resonance Raman process~\cite{Ding2015, Guo2019}, and gradient-echo memory~\cite{Hosseini2011,YWCho2016}, by opening an optical transition pathway between the photons and the magnons via a strong optical control field~\cite{Ou2008Efficient, Hammerer2010QuantumInterface}.

When implementing quantum memories in a quantum network, operations such as entanglement swapping require indistinguishability of the interacting bosonic excitations, which can be characterized by performing Hong-Ou-Mandel interference and observing cancellation of the output state $|1,1\rangle$ in a unitary beam splitter~\cite{HOM1987PRL, Kaufman2014Science, Lopes2015}. This bunching behavior originates from their bosonic nature, i.e., the exchange symmetry of the identical bosons at the beam splitter, while antibunching stems from the Pauli exclusion principle that governs the quantum statistics of identical fermions~\cite{HOMNJP2005}. In the past decades, photon-photon and magnon-magnon quantum interference has been widely demonstrated through various linear beam splitting operations~\cite{HOM1987PRL, Li2016HOMMagnon, Qian2016HOMIndependentSource}, while the indistinguishability between these two distinct excitations has not yet been well verified. Relevant attempts can be found in diamond crystals, in which nonclassical interference between an optical mode and a nonobservable phononic mode has been demonstrated~\cite{England2016}. Recently, new physics on quantum interference has revealed that the bosonic bunching statistics can be switched to antibunching if the beam splitting operation is not unitary, as in the case of an open system with dissipation~\cite{Vest2017Anti,Li2021NonUnitary, vetlugin2021antihongoumandel}. Therefore, a controllable non-Hermitian magnon-photon beam splitting (MPBS) interface~\cite{Wen2019} can lead to the first observation of this exotic physics in cold atomic systems. 

In this Letter, we demonstrate, for the first time, quantum interference between flying photons, from a highly attenuated coherent state, and a single localized ``photon" stored as collective atomic spin excitation~\cite{Duan2001DLCZ,Fleischhauer2005RMP, Gorshkov2007PhotonStorage} in a cold atom ensemble. In our system, the frequency detuning of the control laser field provides a unique tuning knob for the Hermiticity of the coherent magnon-photon interaction. As a result, indistinguishability between the two types of bosons is investigated with varying Hermiticity, showing a transition of quantum statistics from bosonic bunching to fermionlike antibunching. Moreover, the preservation of the quantum interference between these photonic and magnonic modes within the lifetime of the atomic coherence is demonstrated. Therefore, our results establish a new mechanism of interfacing light and matter in a quantum network. 

The principle of the magnon-photon quantum interference is schematically illustrated in Fig.~$\,$\ref{Fig1}(a). On the one hand, by introducing an auxiliary energy level $|4\rangle$, the collective atomic coherence $\hat{S}$ between the states $|1\rangle$ and $|2\rangle$ is produced by a far-detuned pump pulse $\omega_p$ according to the Duan-Lukin-Cirac-Zoller protocol~\cite{Duan2001DLCZ}. A single quantum of collective atomic coherence is generated when a Stokes photon is emitted. On the other hand, the MPBS can be realized through the coherent conversion interaction between the input photons and the magnons, which is stimulated by another control laser $\omega_c$, with a single-photon detuning $\Delta$ from energy level $|3\rangle$~\cite{Wen2019}. Following the experimental procedure in Fig.~$\,$\ref{Fig1}(b), two channels of conversion between magnons and photons occur: the created single atomic coherence can be converted to an anti-Stokes photon with frequency $\omega_{AS}$ or, alternatively, remain stored in the atomic medium; the input optical signal $\omega_{sig}$ can be stored as atomic coherence, while part of it can be still transmitted through the atomic cloud. Since the input optical signal and the readout photons are frequency matched ($\omega_{sig}=\omega_{AS}$) and thus indistinguishable, the two channels of the magnon-photon conversion should also be indistinguishable. Therefore, our scheme allows the investigation of magnon-photon quantum interference, and the experiments are implemented with the setup shown in Fig.~$\,$\ref{Fig1}(c). 


\begin{figure}
\includegraphics[width=1.0\columnwidth]{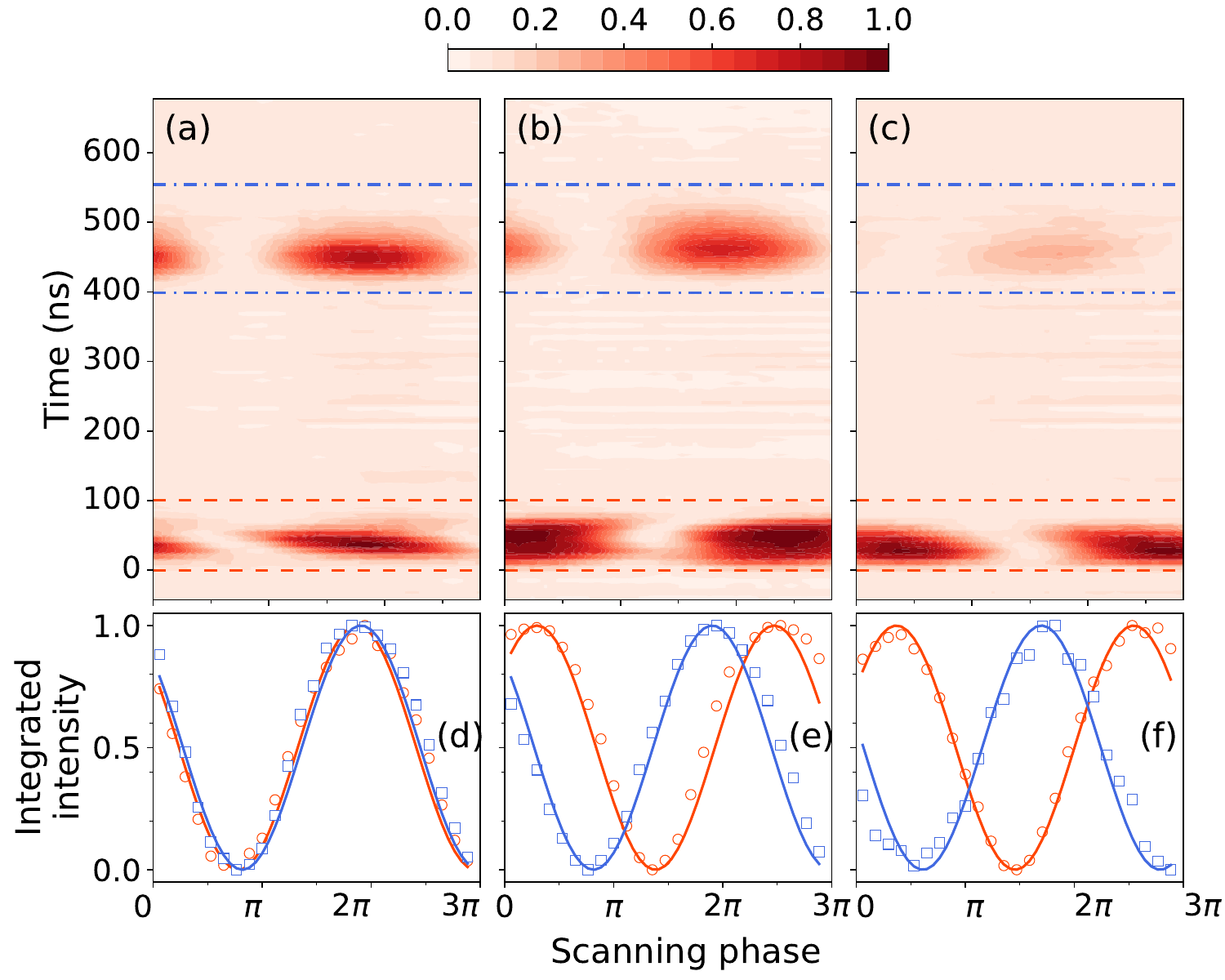} \caption{First-order interference between an optical pulse (dashed lines) and a stored magnon (dot-dashed lines). The normalized intensity of the light and the magnon output with varying interferometric phase determine the phase $\theta_{+}$ of the MPBS. (a) $\theta_{+}=0$ for $\Delta=\SI{0}{\MHz}$, (b) $\theta_{+}=\pi/2$ for $\Delta=\SI{-10}{\MHz}$, and (c) $\theta_{+}=\pi$ for $\Delta=\SI{-50}{MHz}$. The sinusoidal interference curves of the corresponding integrated intensity over respective time window of light (circles) and magnon (squares) pulses are shown in (d)--(f). The solid lines are sinusoidal fits to the experimental data.}
\label{Fig2}
\end{figure}

Theoretically, the evolution of the optical field operator $\hat{a}$ and the coupled atomic spin operator $\hat{S}$ (magnon) obeys the master equations~\cite{supplement} of a typical three-level atomic system, as the one depicted in Fig.~$\,$\ref{Fig1}(a). The essential features of the magnon-photon interactions can be captured by the transformation matrix of a MPBS~\cite{Wen2019,supplement}, which is generally asymmetric and nonunitary, and has the form~\cite{Barnett1998PRA, Uppu2016OE, supplement}
\begin{align}
U_T&=
\left( {\begin{array}{cc}
t & e^{i\theta_{2}}\rho\\
e^{i\theta_{1}}r & \tau
\end{array}} \right),
\label{MPBS matrix}
\end{align}
where $t,r$ and $\tau,\rho$ denote the transmission and reflection coefficients for the photonic and magnonic inputs, respectively, and $\theta_1$($\theta_2$) represents the phase of the reflected (transmitted) output acquired via the conversion of the photonic (magnonic) mode with respect to the input. For an atomic cloud with optical depth $\eta$, and an optical pulse with duration $\tau_{sig} \ll 1/\gamma_{12}$, the phases could be analytically obtained~\cite{supplement} as
\begin{equation}
    \theta_{1} =\arg\left(1-\xi^{-1}\right), \quad
    \theta_{2} =\arg\left(\frac{-\frac{\eta}{\zeta}\left(1-\xi\right)}{1-\frac{\eta}{\zeta}\left(1-\xi\right)}\right),
\label{eq:thelta}
\end{equation}
with the dimensionless parameters $\zeta=\tau_{sig}\Omega_{c}^{2}/{\gamma_{13}}$ and $\xi=\exp{\left[-\zeta/{\left(i\Delta/\gamma_{13}+1\right)}\right]}$. 

Therefore, the phase sum $\theta_{+}=\theta_1+\theta_2$ is critical for determining the Hermiticity of the magnon-photon interaction since the MPBS is non-Hermitian when $\theta_{+}\neq\pi$. It is anticipated that the change of the degree of Hermiticity~\cite{ElGanainy2018NPH} has a dramatic impact on the photon coalescence and gives access to both bunching and antibunching statistics. In the case of a unitary and symmetric beam splitter, $r^2+t^2=1$, only bunching, which corresponds to phase $\theta_{+}=\pi$, can be observed. However, when dissipation, and consequently, the loss of photons and magnons, becomes significant $r^2+t^2\ll1$ and the hybrid system is not anymore locked to a single value of phase $\theta_{+}=\pi$. In a cold atomic ensemble, the degree of dissipation is controlled predominantly by the detuning $\Delta$. The resonant EIT storage scheme ($|\Delta|/\gamma_{13}\approx0$) in this model constitutes a non-Hermitian coupling. Driven by the control beam whose pulse width and power are carefully chosen, the atomic excited state is populated. This allows spontaneous emission to release energy, and the loss channel contributes to the indirect coupling between the coherent magnonic and photonic excitations. By contrast, far-detuned excitation $(|\Delta|/\gamma_{13}\gg1)$ induces storage through Raman transition, and since the population of the excited state is greatly suppressed the loss channel is almost closed.    

Prior to the experiments of quantum interference, we first characterize the MPBS by measuring the first-order interference between the coherent states of magnons and photons. The interference is realized by injecting two sequential pulses with different relative phases to the cold atomic ensemble, with the first one generating the coherent magnon excitations via EIT~\cite{Wen2019} while the second one serving as the coherent signal input. By scanning the relative phase between these two pulses, the temporal dynamics of the pulsed light and magnon outputs for different control laser detuning $\Delta$ are recorded, and their intensities normalized to the highest value are plotted in Figs.~\ref{Fig2}(a)--\ref{Fig2}(c). We note that the coherent magnon output can be optically detected after its conversion to an anti-Stokes pulse by the control beam. From the normalized integrated intensities, as plotted in Figs.~\ref{Fig2}(d)--\ref{Fig2}(f), the sinusoidal interference curves show first-order interference and the phase $\theta_{+}$ is extracted. For resonant control laser ($|\Delta|/\gamma_{13}\approx0$), as shown in Figs.~\ref{Fig2}(a) and (d), the intensities of both photon and magnon outputs exhibit correlation in agreement with the theoretical prediction for $\theta_{+}=0$~\cite{supplement}. By contrast, for the far-detuned case with $|\Delta|/\gamma_{13}\gg1$ [Figs.~\ref{Fig2}(c) and (f)], the outputs exhibit complementary change for a varying phase, which indicates $\theta_{+}=\pi$. By controlling the detuning $\Delta$ of the MPBS we can also access other values of $\theta_{+}$ between the two extreme cases $0$ and $\pi$, such as $\theta_{+}=\pi/2$ as shown in Figs.~\ref{Fig2}(b) and (e).

Consequently, quantum interference between photons and single quanta of atomic coherence is experimentally studied by interfering a weak photonic coherent input $|\alpha\rangle_{a}$ and a heralded single-magnon state $|\psi\rangle_{b}$. The quantum behavior in the MPBS can be characterized by the normalized second-order correlation $g^{\left(2\right)}(0)$. $g^{\left(2\right)}(0)$ is essentially the normalized coincidence in a Hong-Ou-Mandel interferometer and would be 0 for quantum interference of identical single bosons and 2 for identical single fermions in an ideal 50:50 beam splitter. To account for any imperfections in its preparation the magnon takes the form of a nonideal Fock state $|\psi\rangle_{b}=\sqrt{p_{0}}|0\rangle+\sqrt{p_{1}}|1\rangle+\sqrt{p_{2}}|2\rangle$, where $p_n$ is the probability for an n-magnon state, and the expected $g^{(2)}(0)$ is expressed as~\cite{supplement, Qian2016HOMIndependentSource,Duan2020HOMSunPhoton} 
\begin{widetext}
\begin{equation}
    g^{\left(2\right)}(0)=1+\frac{2\cos(\theta_{+})rt\tau\rho H_{ab}(0)}{t^{2}r^{2}|\alpha|^{2}/p_{1}+\rho^{2}\tau^{2}g^{(2)}_{m}(0)p_{1}/|\alpha|^{2}+(t^{2}\tau^{2}+r^{2}\rho^{2})H_{ab}(0)}.
\label{gfactor}
\end{equation}
\end{widetext}
Here, $H_{ab}(0)=\int dt_{0}|h_{a}(t_{0})h_{b}(t_{0})|^{2}$ is a measure of the temporal overlap between the input wave packets $h_{a}$ and $h_{b}$~\cite{supplement}. In our experiments, $g^{\left(2\right)}(0)$ is obtained as the ratio between the coincidence counts at $\Delta t=0$, where photons and magnons completely interfere, divided by the coincidence counts at $\Delta t=\pm T$ when they are completely distinguishable in time~\cite{supplement}. The normalized second-order self-correlation of the magnon is expected to be $g_{m}^{(2)}(0)\approx 2p_{2}/p_{1}^{2}$. The criterion $g_{m}^{(2)}(0)<0.5$ for single-magnon input is satisfied for all the experiments reported here.

\begin{figure}
\includegraphics[width=1.0\columnwidth]{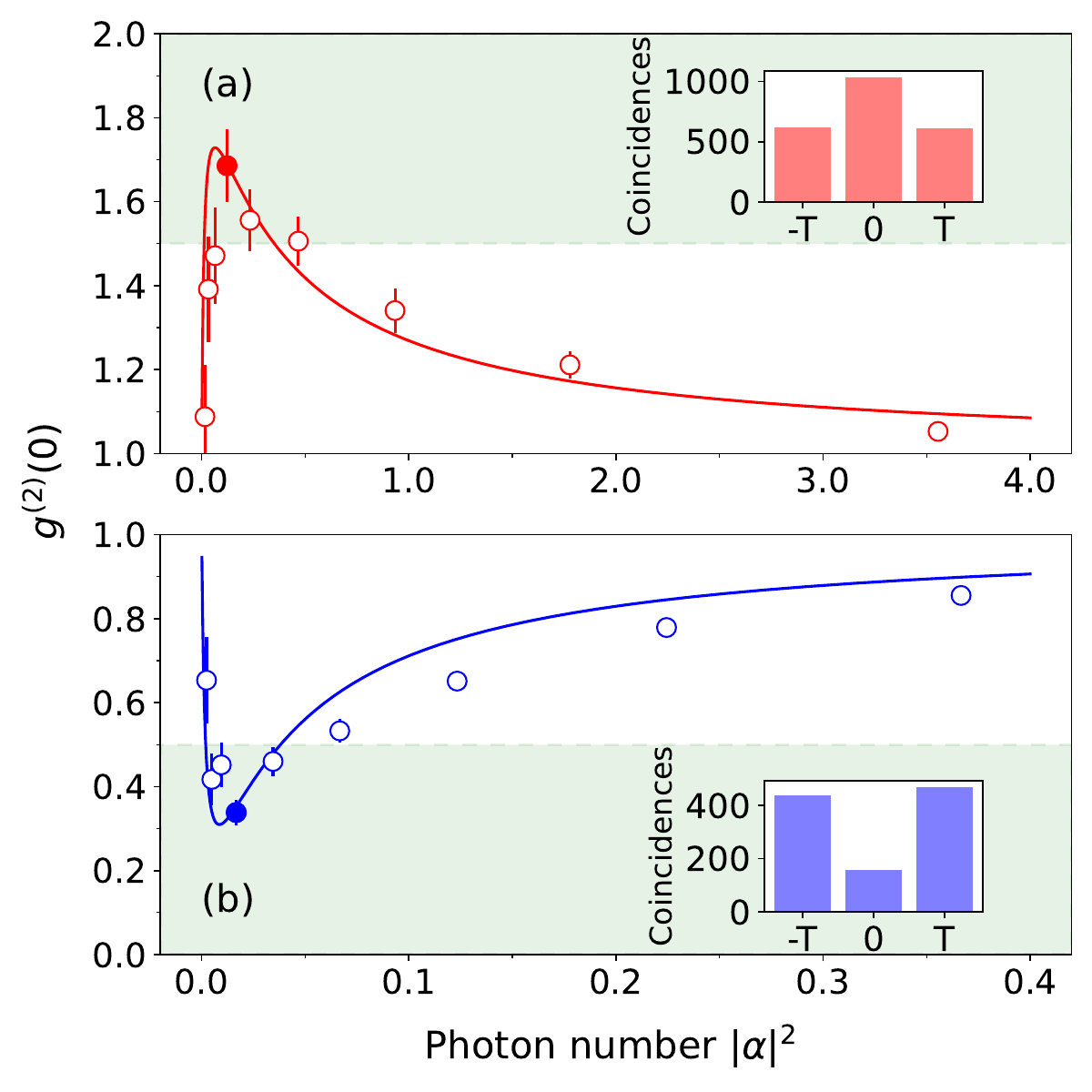} 
\caption{The $g^{\left(2\right)}(0)$ of quantum interference versus average photon number per coherent signal pulse for (a) $\theta_{+}=0$ with fermionlike antibunching and (b) $\theta_{+}=\pi$ with bosonlike bunching. The shaded regions mark the correlation values that are classically forbidden. In the insets are the total coincidence counts used to calculate the $g^{\left(2\right)}(0)$ for the intensities of the highest violation of the classical limits (the solid dot in each panel). The solid lines are the theory predictions of Eq.$\,$\ref{gfactor}. The error bars are deduced by assuming Poissonian distribution of the coincidence counts.}
\label{Fig3}
\end{figure}

Figure~\ref{Fig3} demonstrates the transition from bosonic bunching to fermionlike antibunching behavior for the quantum interference between photons and stored magnons. By varying the average input photon number $|\alpha|^2$, statistically significant violations of the classical limits for the $g^{\left(2\right)}(0)$ are observed for both the antibunching [Fig.~\ref{Fig3}(a)] and the bunching cases [Fig.~\ref{Fig3}(b)]. Because of the balance between the suppression of multiphoton and multimagnon events, optimal violations are achieved around $|\alpha|^2=\sqrt{2p_2}(\tau\rho /tr)=0.064$ for the on-resonance case, and 0.0086 for the far off-resonance case, respectively. The highest observed value of the quantum interference visibility $V=\left|1-g^{(2)}(0)\right|$ is larger than $0.5$ for both cases $0.69~(0.67)$. However, with much lower photon intensities, the accidental two-magnon events stemming from the nonzero $g_{m}^{(2)}(0)$ dominate and the interference at the MPBS is washed out. Similarly, the excess of multiphoton events limits the visibility for larger $|\alpha|^2$. Another relevant limiting factor could be asymmetry of the transmission or reflection coefficients~\cite{supplement} and special care was taken to reduce it. Our experimental observations are in quantitative agreement with theoretical predictions, from Eq.$\,$\ref{gfactor}, for both cases with phases $\theta_{+}=0$ (antibunching) and $\theta_{+}=\pi$ (bunching). The transmission and reflection coefficients for the case of on(off)-resonance $t^{2}=0.025~(0.489)$, $r^{2}=0.030~(0.108)$, $\tau^{2}=0.026~(0.017)$, $\rho^{2}=0.035~(0.045)$, and $g_{m}^{(2)}(0)=0.118~(0.170)$ that characterizes the single-magnon input, have been extracted independently by utilizing our MPBS. By considering the overall photon collection efficiency of about 0.36 the corresponding single-magnon probability becomes $p_{1}=0.167~(0.172)$. In the theoretical curves of Fig.~\ref{Fig3}, the temporal wave packet overlap is set to $H_{ab}(0)=1$, since even for the lowest observed value of 0.89 its influence on the $g^{\left(2\right)}(0)$ is negligible~\cite{supplement}.

\begin{figure}
\includegraphics[width=1.0\columnwidth]{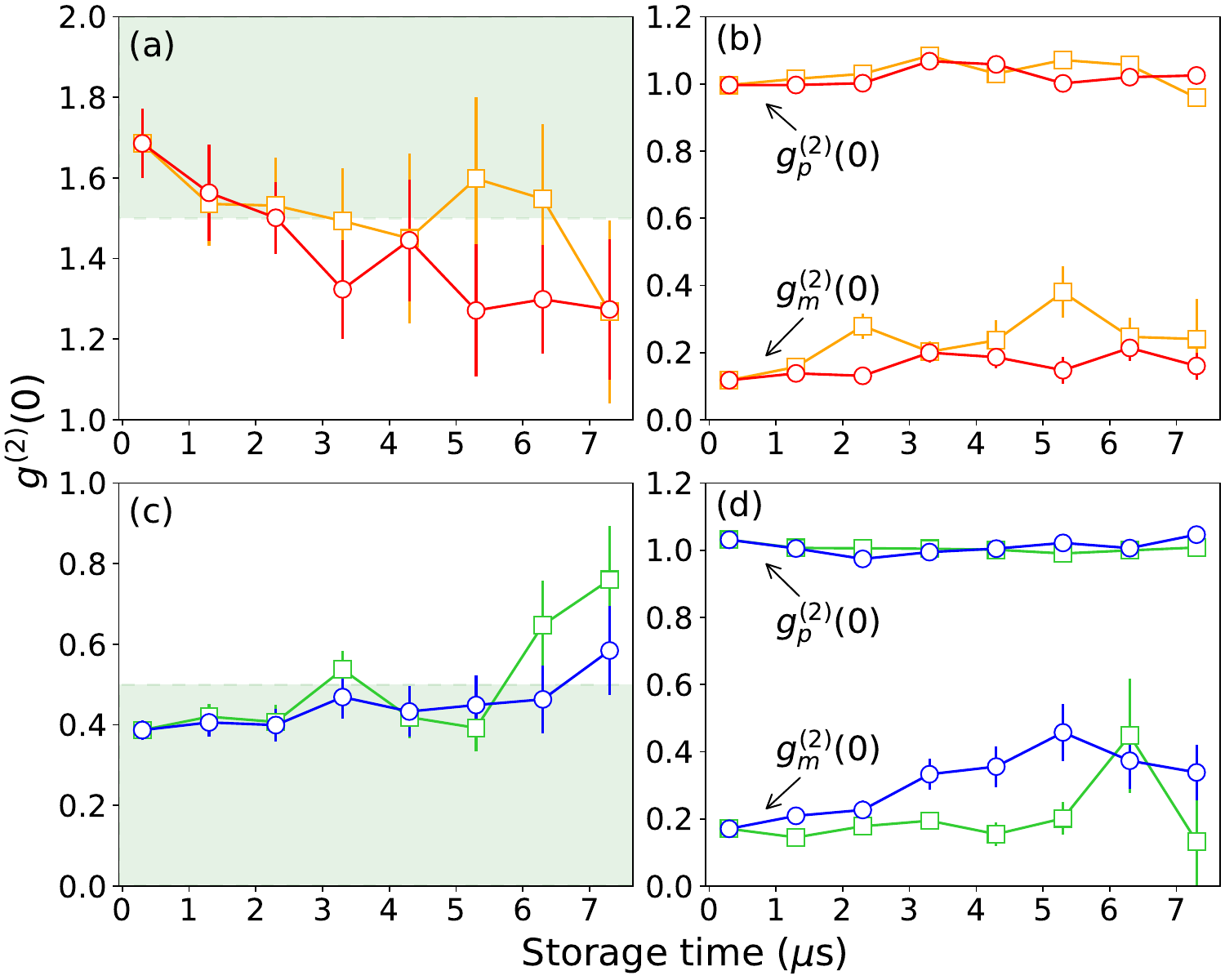} 
\caption{The $g^{\left(2\right)}(0)$ of quantum interference versus storage time of the collective atomic coherence for (a) $\theta_{+}=0$ and (c) $\theta_{+}=\pi$ that lead respectively to magnon-photon antibunching and bunching. The magnon storage times before(after) the beam splitting action are marked with squares(circles). The corresponding second-order self-correlations $g^{(2)}_p(0)$ for photons and $g^{(2)}_m(0)$ for magnons are shown in (b) and (d). The shaded regions mark the correlation values that are classically forbidden. The error bars are deduced by assuming Poissonian distribution of the coincidence counts.}
\label{Fig4}
\end{figure}

To investigate whether the indistinguishability between stored single magnons and single-photon-level optical pulses can be preserved, we further study the temporal evolution of second-order correlation functions against the storage time for magnon-photon, photon-photon, and magnon-magnon correlations. Here, we choose the points with the highest violation of the classical limit in Fig.~\ref{Fig3} and compare the correlation functions for the output before or after the operation of the MPBS, following the experimental sequence shown in Fig.\,\ref{Fig1}(b). The coherent pulse storage lifetime is optimized to be longer than $10\,\mathrm{\mu s}$, so this would not be the main limiting factor for the decay of $g^{\left(2\right)}(0)$. We measure the resilience of the quantum interference in both cases as shown in Figs.~\ref{Fig4}(a) and \ref{Fig4}(c). We found that $g^{\left(2\right)}(0)$ approaches the classical limit when the storage time exceeds $3~(6)\,\mathrm{\mu s}$ for the on-resonance (far-detuned) configurations. The decay of $g^{\left(2\right)}(0)$ toward the classical limit $1.5~(0.5)$ is mainly attributed to the variation of the magnons due to the ground-state dephasing $\gamma_{12}$. We expect that improvement of the magnetic field stability and further cooling of the atomic cloud can address this issue. A second cause is that the multiphoton noise starts to play a more significant role when the storage time increases and therefore the retrieved photon counts are reduced to the background noise. This can also be seen in Figs.~\ref{Fig4}(b) and \ref{Fig4}(d) illustrating the second-order self-correlation of the magnon $g^{(2)}_{m}(0)$, which increases with time and whose error bars become larger. By contrast, the photonic coherent state is impervious to attenuation and noise, and thus its second-order self-correlation $g^{(2)}_{p}(0)\approx1$ is preserved for all storage times. 

In summary, in this magnon-photon conversion process, we observe the interference of photons and single quanta of atomic coherence entering the quantum regime. When we tune the phase sum in the beam splitter matrix from $\pi$ to $0$, bosonic bunching can be switched to antibunching which is commonly associated with fermions. The Hermiticity of the atom-light beam-splitter-type interface can be controlled by frequency detuning the storage scheme. The physics of this switching mechanism is fundamentally different from earlier studies exploiting nonlinear or polarization entangled states ~\cite{Lahini2010, Matthews2013, N00NPRL2016}. The switching speed between bosonlike and fermionlike statistics is promisingly high since the laser frequency can change in just a few nanoseconds. The demonstrated magnon-photon interface may enable a hybrid quantum network system relying on photons and localized atomic excitations. The controllable Hermiticity of the beam splitter can lead to magnon-photon distributions different from those occurring in boson sampling. Exciting possibilities include quantum interference of photonic and atomic Fock states, multiport nonunitary hybrid beam splitters, engineering non-Hermitian temporal Floquet Hamiltonians, $\mathcal{PT}$-symmetric quantum optics~\cite{ElGanainy2018NPH, Klauck2019}, and creation of hyperentangled states with polarization degrees of freedom. The concepts demonstrated here in a cold atomic ensemble could be expanded to a number of other physical systems that involve elementary excitation storage such as diamond vacancy centers~\cite{England2016}, trapped ions, rare-earth ions in solids, hot atomic vapors, and even optical microresonators. 

\begin{acknowledgments}
We thank Professor Weiping Zhang for helpful discussions. This work is supported by the Guangdong Provincial Key Laboratory (Grant No. 2019B121203002) and the National Natural Science Foundation of China (NSFC) through Grants No. 12074171 and No. 12074168. J. F. C. acknowledges the Guangdong Key Project under Grant No. 2019ZT08X324, and the Guangdong Innovation Project under Grant No. 2019KTSCX160. C.-L. Z. is supported by NSFC Grants No. 11922411 and No. 11874342. R. W. is supported by NSFC Grant No. 12104334.
\end{acknowledgments}

\bibliography{InterferencePhotonMagnon} 

\pagebreak 

\clearpage

\widetext

\appendix

\renewcommand{\appendixname}{Supplemental Material}
\renewcommand{\thefigure}{S\arabic{figure}}
\setcounter{figure}{0}
\setcounter{equation}{0}

\counterwithout{equation}{section}
\renewcommand{\theequation}{S\arabic{equation}}

\section{Experimental methods}
\begin{figure*}[h]
\includegraphics[width=0.5\textwidth]{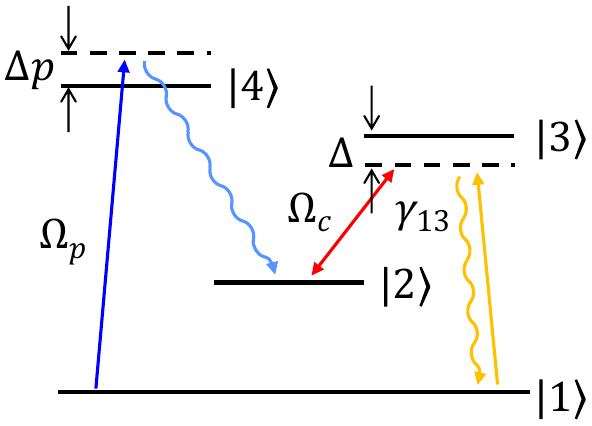} \caption{Magnon-photon conversion channels of the double-$\Lambda$ configuration of $^{85}$Rb. Ground hyperfine states $|1\rangle~(|2\rangle)$: $5S_{1/2}, F=2~(F=3)$; and the two excited states $|3\rangle$: $5P_{1/2},F=3$  and $|4\rangle$: $5P_{3/2},F=3$ for the magnon-photon conversion and the magnon creation channel, respectively.}
\label{SMFig1}
\end{figure*}

A four-level system in a double-$\Lambda$ configuration, as shown in Fig.~\ref{SMFig1}, is used for our quantum interference experiments between single magnons and photons. We hereby further clarify the terms of ``reflection" and ``transmission" for these two types of inputs in our magnon-photon beam splitter. Magnons are created by the DLCZ protocol~\cite{Duan2001DLCZ} as follows: a blue-detuned pump light pulse excites atoms from state $|1\rangle$ to $|4\rangle$, and then spontaneous decay to state $|2\rangle$ creates a single magnon, which is heralded by the emission of a single Stokes photon. The conversion process from magnon into photon with a strong control laser pulse is called ``reflection" for the magnon. The same laser is used to read out the stored magnon by exciting any atoms that are in state $|2\rangle$ to state $|3\rangle$. The subsequent spontaneous emission of a single photon when returning back to state $|1\rangle$ signals the single stored magnon, and it is the ``transmission" of the magnon. Any undetected magnons that are still left in the system are read out by another control pulse, which has the additional function of system initialization. The input photons, on the other arm of the beam splitter, that enter the atomic cloud come from the simultaneously shining signal light pulse and control light pulse in an electromagnetically induced transparency (EIT) scheme. The control light opens a transmission window through state $|1\rangle$ to state $|3\rangle$ for the signal light and this is called the ``transmission" channel for the photons. The photons that are converted into magnons are read out as above and this is the photon ``reflection" channel.\par
A cigar-shaped cylindrical atomic cloud of $^{85}$Rb is laser cooled to \SI{120}{\micro\kelvin} in a two-dimensional magneto-optical trap (MOT). The optical depth (OD) of the atomic cloud can be tuned from 10 to 50 by varying the power of a repump light (not shown here). The pump laser used for the DLCZ protocol is blue-detuned by $\Delta_{p}=\SI{120}{\MHz}$ with respect to the transition $|1\rangle$: $5S_{1/2},F=2$ to $|4\rangle$: $5P_{3/2},F=3$. It is frequency shifted  by an acousto-optical modulator (AOM) and its pulse duration is approximately $\tau_{p}=\SI{200}{\ns}$ with a peak value of Rabi frequency $\Omega_{p}$ around $2\gamma_{13}$, where $\gamma_{13}=2\pi\cdot\SI{3}{MHz}$ is the decay rate from the excited state $|3\rangle$ to the ground state $|1\rangle$. The pump and the control laser beams are counter-propagating and are collimated with an $e^{-2}$ radius of $\SI{0.6}{mm}$, which is enough to cover the whole atomic cloud. The signal(probe) and control light are coherent with each other since they come from the same laser. Both have a single-photon detuning $\Delta$, introduced by AOMs, from $|1\rangle$: $5S_{1/2},F=2$ and $|2\rangle$: $5S_{1/2},F=3$ to $|3\rangle$: $5P_{1/2},F=3$, respectively, while remaining at all times on a two-photon resonance. Two amplitude electro-optical modulators (EOM) modify the pulse profile of both the control and probe light. The peak value of the control light Rabi frequency $\Omega_{c}$ ranges from $\gamma_{13}$ to $5\gamma_{13}$, in order to get appropriate $\zeta=|\Omega_{c}|^{2}\tau_{sig}/\gamma_{13}$ (see section \ref{SMSec2}) for the different single-photon detuning $\Delta$ cases. The optical signal beam is focused to a radius of $\SI{125}{\um}$ at the center of the MOT and is collected on the opposite side. We initially observe the first-order interference phase similarly to Ref.~\cite{Wen2019}, the difference being that for our experiments the relative phase of light and spin wave is modulated directly by a phase EOM in the probe light path. By tuning the OD and the single-photon detuning $\Delta$, we can set the sum phase of the beam splitter to be 0, $\pi/2$ and $\pi$, with parameters OD $=10$, $\Delta=\SI{0}{\MHz}$, OD $=30$, $\Delta=\SI{-10}{\MHz}$, and OD $=30$, $\Delta=\SI{-50}{\MHz}$, as depicted in Fig. 2 of the main text. \par
Two-particle interference demands to reach single quantum level for both photons and magnons. The magnons are created by the pump pulse using the DLCZ protocol as previously described. We emphasize here that the photon input is an extremely weak, attenuated coherent probe light at the single photon level. The emitted Stokes and read-out anti-Stokes photons are collected by single-mode fibers with an efficiency of $90\%$, then sent to the single photon detector modules (SPCM) with a detection efficiency of $50\%$. Narrow band ($\text{FWHM}=\SI{500}{\MHz}$) temperature stabilized etalons were used to block stray photons outside of the desired frequency range, with transmission of $80\%$ both for the Stokes and the anti-Stokes photons, thus getting an overall collection efficiency of $36\%$ as mentioned in the main text. \par
The photon (magnon) transmission $t^2~(\tau^2)$ and reflection $r^2~(\rho^2)$ coefficients (see section \ref{SMSec3}) have been determined experimentally. The coefficients $t^2~(r^2)$ for a single-port photon input are given by the ratio of the transmitted (stored) photons over the transmitted photons when no atoms were present. Similarly, the coefficients $\tau^2~(\rho^2)$ for a single-port magnon input are given by the ratio of the stored (transmitted) photons over the number of the Stokes photons that herald the magnon generation. \par
\begin{figure*}[h]
\includegraphics[width=0.5\textwidth]{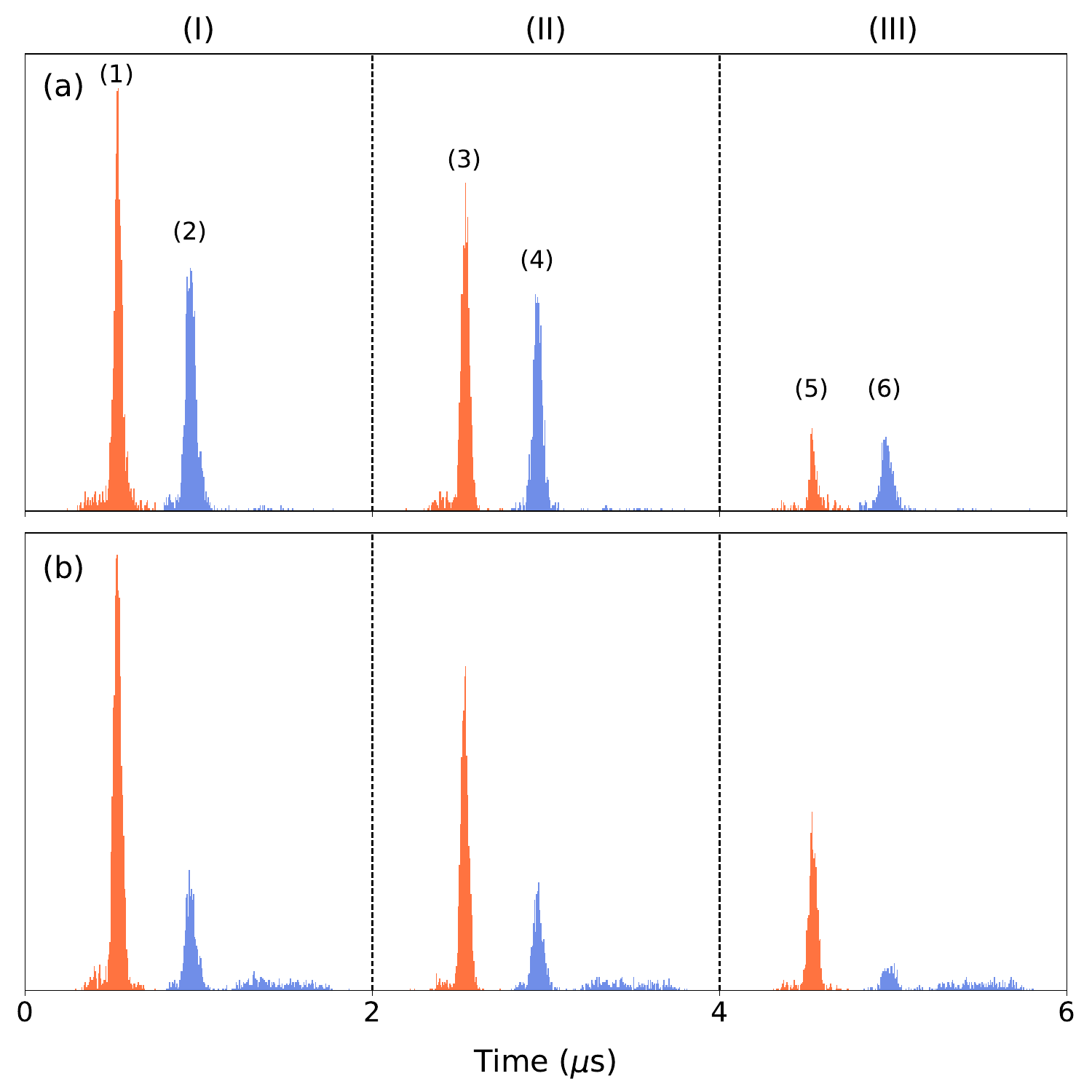} 
\caption{Typical output of the beam splitter detected by SPCM. (a) and (b) demonstrate the case of $\theta_{+}=0$ and $\pi$, respectively. Output for part (I) one input port being a single magnon, and the other one a single photon level coherent state light, and part (II) only one port having a single magnon input. The orange peak (3) and the blue peak (4) are the reflection and transmission of magnons. Finally, part (III) is the output of the beam splitter with only one port single photon level coherent state light input. The orange peak (5) and the blue peak (6) are now the transmission and reflection of photons. Swapping the order of the parts (II) and (III) leads to the times $\Delta t=\pm T$ of the insets in Fig. 3 of the main text.}
\label{SMFig2}
\end{figure*}

The normalized second-order correlation $g^{(2)}(0)$ is calculated by dividing the coincidence between the transmission and the reflection ((1)--(2) in Fig.~\ref{SMFig2}) of the beam splitter when both ports (magnon and photon) have an input, with the sum of self- ((3)--(4), (5)--(6) in Fig. \ref{SMFig2}) and cross- ((3)--(6), (4)--(5) in Fig.~\ref{SMFig2}) coincidence between the two outputs when only one port has an input (magnon or photon). We can also calculate from the same measurements the second-order self-correlation $g^{(2)}(0)$ of each input source by counting the coincidence between transmission and reflection
\begin{equation}
    g^{(2)}(0)=\frac{N_{GTR}N_{G}}{N_{GT}N_{GR}},
\end{equation}
where $N_{GTR}$, $N_{G}$, $N_{GT}$, and $N_{GR}$ are respectively the number of gated coincidences between transmission and reflection, gate itself, gated transmission, and gated reflection. \par 
We perform the quantum intereference experiments by using a time sequence with a cycle period of \SI{6}{\us} that consists of three separate beam splitting actions, as depicted in Fig.~\ref{SMFig2}: part (I) for a two-port input, part (II) only for a magnon input, and part (III) only for a photon input independently. Atoms are being loaded in the MOT with \SI{100}{\hertz} repetition rate and the measurement has a $10\%$ duty cycle, so in our \SI{1}{\ms} experiment window the quantum interference sequence is repeated 160 times. The shortest total accumulation times for the case of sum phase 0 and $\pi$, in Fig. 3 of the main text, are 1 hours and 0.5 hours, respectively. With the average photon number being reduced, they increase to 7 hours and 1 hours. In the experiment of indistinguishability storage, the extension of storage time increases consequently the cycle period, and thus the measurement of each data point in Fig. 4 of the main text requires at least 5 hours. \par
To demonstrate the ability of indistinguishability storage, we optimized the dephasing rate $\gamma_{12}$ by turning off the trapping magnetic field of the MOT and using extra coils to compensate earth's magnetic field, and any additional stray fields. Using conventional EIT storage protocol, the measured $e^{-1}$ storage lifetimes for on-resonance and far-resonance are \SI{15}{\us} and \SI{14}{\us}.  

\section{Heisenberg Master equation description}\label{SMSec2}

From the Hamiltonian that describes how an input optical field interacts with an ensemble of cold atoms driven by a strong coherent field with Rabi frequency $\Omega_{c}$ coupled in a three-level configuration, when both fields have a single-photon detuning $\Delta$ but maintain two-photon resonance, we get the coupled equations of motion~\cite{Hammerer2010QuantumInterface,Gorshkov2007PhotonStorage} 
\begin{equation}
\begin{aligned}
    &c\frac{\partial}{\partial z}\mathcal{E}=-\frac{g^{2}N}{i\Delta+\gamma_{13}}\mathcal{E}-\frac{g\sqrt{N}\Omega_{c}}{i\Delta+\gamma_{13}}S,\\
    &\frac{\partial}{\partial t}S=\left(-\gamma_{12}-\frac{|\Omega_{c}|^{2}}{i\Delta+\gamma_{13}}\right)S-\frac{g\sqrt{N}\Omega^{*}_{c}}{i\Delta+\gamma_{13}}\mathcal{E},
\end{aligned}
\end{equation}
where $\mathcal{E}=\sum_{k}a_{k}$ is the slowly varying operator of the input field on the mode $a_{k}$ base, and $S=\sqrt{N}\sigma_{12}$ is the spin wave operator with a slowly varying atom operator $\sigma_{12}=1/N\sum_{j}\sigma_{12}^{j}(z_{j},t)$ that includes all the atoms in $z$ direction. $g=\mu_{13}\sqrt{\omega_{sig}/2\hbar\varepsilon_{0}V}$ is the atom-field coupling constant of the input field, $\mu_{13}$ is the dipole moment of the transition $|1\rangle\leftrightarrow|3\rangle$, and  $V$ is the quantization volume. $N$ is the number of atoms in this area, and $L$ is the length of the medium in $z$ direction. $\gamma_{\mu\nu}$ is the nonzero decay from state $\nu$ to state $\mu$ due to spontaneous emission.\par
We introduce a new scaled field 
\begin{equation}
    E=\frac{g\sqrt{N}}{\Omega_{c}}\mathcal{E},
\end{equation}
and then the equations of motion become
\begin{equation}
\begin{aligned}
    &c\frac{\partial}{\partial z}E=-\frac{g^{2}N}{i\Delta+\gamma_{13}}E-\frac{g^{2}N}{i\Delta+\gamma_{13}}S,\\
    &\frac{\partial}{\partial t}S=\left(-\gamma_{12}-\frac{|\Omega_{c}|^{2}}{i\Delta+\gamma_{13}}\right)S-\frac{|\Omega_{c}|^{2}}{i\Delta+\gamma_{13}}E.
\end{aligned}
\end{equation}\par
If we treat $S$ as a single uniform mode $m$ in a thin atomic medium, and ignore the $\gamma_{12}$ for pulse widths $\tau_{sig}\ll 1/\gamma_{12}$~\cite{Wen2019}
\begin{equation}
\begin{aligned}
    &E_{out}=\left(1+\frac{\eta\gamma_{13}}{-i\Delta-\gamma_{13}}e^{\frac{|\Omega_{c}|^{2}}{-i\Delta-\gamma_{13}}t}\right)E_{in}+\frac{\eta\gamma_{13}}{-i\Delta-\gamma_{13}}e^{\frac{|\Omega_{c}|^{2}}{-i\Delta-\gamma_{13}}t}m(0),\\
    &m(t)=e^{\frac{|\Omega_{c}|^{2}}{-i\Delta-\gamma_{13}}t}m(0)-\left(1-e^{\frac{|\Omega_{c}|^{2}}{-i\Delta-\gamma_{13}}t}\right)E_{in},
\end{aligned}
\end{equation}
where $\eta=g^{2}NL/c\gamma_{13}$ is the OD of the atomic medium.

The input-output relation of the magnon-photon beam splitter is
\begin{equation}
    \left(\begin{array}{c}
    \mathscr{E}_{out} \\
    M_{out}
    \end{array}\right)=\left(\begin{array}{cc}
    1-\frac{\eta \gamma_{13}}{|\Omega_{c}|^{2} \tau_{sig}}\left(1-e^{\frac{|\Omega_{c}|^{2}}{-i \Delta-\gamma_{13}} \tau_{sig}}\right) & -\frac{\eta \gamma_{13}}{|\Omega_{c}|^{2} \tau_{sig}}\left(1-e^{\frac{|\Omega_{c}|^{2}}{-i \Delta-\gamma_{13}} \tau_{sig}}\right)\\
    -\left(1-e^{\frac{|\Omega_{c}|^{2}}{-i \Delta-\gamma_{13}} \tau_{sig}}\right) & e^{\frac{|\Omega_{c}|^{2}}{-i \Delta-\gamma_{13}} \tau_{sig}}
    \end{array}\right)\left(\begin{array}{c}
    \mathscr{E}_{in} \\
    M_{in}
    \end{array}\right),
\end{equation}
where $M_{in}=m(0)$, $M_{out}=m(\tau_{sig})$, and $\mathscr{E}_{in(out)}=\int_{0}^{\tau_{sig}}E_{in(out)}dt/\tau_{sig}$.

By introducing the dimensionless parameter $\zeta=|\Omega_{c}|^{2}\tau_{sig}/\gamma_{13}$, we have
\begin{equation}
\begin{aligned}
    \left(\begin{array}{c}{\mathscr{E}_{out}} \\ {M_{out}}\end{array}\right)&=\left(\begin{array}{cc}
    1-\frac{\eta}{\zeta}\left(1-e^{-\frac{\zeta}{i \Delta / \gamma_{13}+1}}\right) & -\frac{\eta}{\zeta}\left(1-e^{-\frac{\zeta}{i \Delta / \gamma_{13}+1}}\right)\\
    -\left(1-e^{-\frac{\zeta}{i \Delta / \gamma_{13}+1}}\right) & e^{-\frac{\zeta}{i \Delta / \gamma_{13}+1}}
    \end{array}\right)\left(\begin{array}{c}{\mathscr{E}_{in}} \\ {M_{in}}\end{array}\right)\\
    &=\left(\begin{array}{cc}
    1-\frac{\eta}{\zeta}\left(1-\xi\right) & -\frac{\eta}{\zeta}\left(1-\xi\right)\\
    -\left(1-\xi\right) & \xi
    \end{array}\right)\left(\begin{array}{c}{\mathscr{E}_{in}} \\ {M_{in}}\end{array}\right),
\end{aligned}
\end{equation}
with $\xi=e^{-\frac{\zeta}{i\Delta/\gamma_{13}+1}}$, and the conversion phases
\begin{equation}
\begin{aligned}
    \theta_{1} &=\arg\left(1-\xi^{-1}\right), \\
    \theta_{2} &=\arg\left(\frac{-\frac{\eta}{\zeta}\left(1-\xi\right)}{1-\frac{\eta}{\zeta}\left(1-\xi\right)}\right).
\end{aligned}
\end{equation}

\section{Quantum optics of an asymmetric and unbalanced beam splitter}\label{SMSec3}

\subsection{Basic description of an asymmetric beam splitter}

\begin{figure*}[h]
\includegraphics[width=0.5\textwidth]{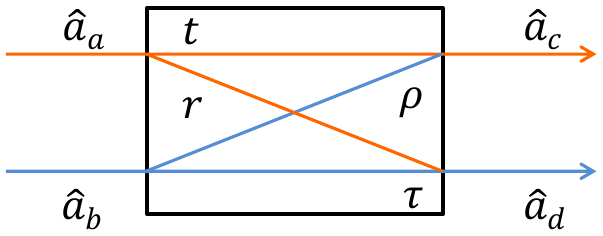} 
\caption{Schematic diagram of an asymmetric beam splitter. The input ports are $\hat{a}_{a}$ and $\hat{a}_{b}$, and the output ports are $\hat{a}_{c}$ and $\hat{a}_{d}$. The transmission and reflection coefficients for each input port are $t-r$ and $\tau-\rho$, respectively.}
\label{SMFig3}
\end{figure*}

We note that, the results of this section are valid for both photon-photon beam splitters as well for magnon-photon ones. For clarity purposes, we refer to ports $b$ and $d$ as the ``magnon" ports.

For an asymmetric beam splitter~\cite{Uppu2016OE} the reflection and transmission coefficients are not necessarily equal and therefore we define the general transformation matrix as  
\begin{equation}
   U_{T}=\left(\begin{array}{cc}{t}&{e^{i\theta_{2}}\rho}\\{e^{i\theta_{1}}r}&{\tau}\end{array}\right). 
\end{equation}
We assume that the beam splitter is a broadband device and therefore the field operator is independent of the photon frequency. This is a good approximation for the narrow linewidth photon source generated from atomic ensembles. Thus, the field operator at the two outputs $c$ and $d$ can be expressed in the form of~\cite{Legero2003time} 
\begin{equation}
\begin{aligned}
    &\mathbf{E}_{c}^{+}(t^{\prime})=h_{c}(t^{\prime})\hat{a}_{c}=t h_{a}(t^{\prime})\hat{a}_{a}+e^{i\theta_{2}}\rho h_{b}(t^{\prime})\hat{a}_{b},\\
    &\mathbf{E}_{d}^{+}(t^{\prime})=h_{d}(t^{\prime})\hat{a}_{d}=e^{i\theta_{1}}r h_{a}(t^{\prime})\hat{a}_{a}+\tau h_{b}(t^{\prime})\hat{a}_{b},
\end{aligned}
\end{equation}
where $h_{i}(t^{\prime}), (i=a,b,c,d)$ are the normalized shapes of the temporal wave packets. In the case of a probabilistic single-photon source obtained from paired photons the time $t^{\prime}$ denotes the relative time with respect to the arrival of the trigger photons.\par

The joint probability of detection events in the two output ports of the beam splitter at times $t_{0}$ and $t_{0}+\Delta t$ is given by 
\begin{equation}
\begin{aligned}
    P(t_{0}, \Delta t)=&\langle\mathbf{E}_{c}^{-}(t_{0})\mathbf{E}_{d}^{-}(t_{0}+\Delta t)\mathbf{E}_{d}^{+}(t_{0}+\Delta t)\mathbf{E}_{c}^{+}(t_{0})\rangle\\
    =&\langle(t h_{a}^{*}(t_{0})\hat{a}_{a}^{\dagger}+e^{-i\theta_{2}}\rho h_{b}^{*}(t_{0})\hat{a}_{b}^{\dagger})(e^{-i\theta_{1}}r h_{a}^{*}(t_{0}+\Delta t)\hat{a}_{a}^{\dagger}+\tau h_{b}^{*}(t_{0}+\Delta t)\hat{a}_{b}^{\dagger})\\
    &\cdot(e^{i\theta_{1}}r h_{a}(t_{0}+\Delta t)\hat{a}_{a}+\tau h_{b}(t_{0}+\Delta t)\hat{a}_{b})(t h_{a}(t_{0})\hat{a}_{a}+e^{i\theta_{2}}\rho h_{b}(t_{0})\hat{a}_{b})\rangle.\\
\end{aligned}
\end{equation}
The expansion of equation above has 16 items, while only 6 terms give a contribution 
\begin{equation}
\begin{aligned}
    P(t_{0}, \Delta t)&=t^{2}r^{2}|h_{a}(t_{0})h_{a}(t_{0}+\Delta t)|^{2}\langle\hat{a}_{a}^{\dagger}\hat{a}_{a}^{\dagger}\hat{a}_{a}\hat{a}_{a}\rangle\\
    &+\rho^{2}\tau^{2}|h_{b}(t_{0})h_{b}(t_{0}+\Delta t)|^{2}\langle\hat{a}_{b}^{\dagger}\hat{a}_{b}^{\dagger}\hat{a}_{b}\hat{a}_{b}\rangle\\
    &+t^{2}\tau^{2}|h_{a}(t_{0})h_{b}(t_{0}+\Delta t)|^{2}\langle\hat{a}_{a}^{\dagger}\hat{a}_{b}^{\dagger}\hat{a}_{b}\hat{a}_{a}\rangle\\
    &+r^{2}\rho^{2}|h_{a}(t_{0}+\Delta t)h_{b}(t_{0})|^{2}\langle\hat{a}_{b}^{\dagger}\hat{a}_{a}^{\dagger}\hat{a}_{a}\hat{a}_{b}\rangle\\
    &+e^{-i(\theta_{1}+\theta_{2})}\rho r\tau t h_{b}^{*}(t_{0})h_{a}^{*}(t_{0}+\Delta t)h_{p}(t_{0}+\Delta t)h_{a}(t_{0})\langle\hat{a}_{b}^{\dagger}\hat{a}_{a}^{\dagger}\hat{a}_{b}\hat{a}_{a}\rangle\\
    &+e^{i(\theta_{1}+\theta_{2})}t\tau r\rho h_{a}^{*}(t_{0})h_{b}^{*}(t_{0}+\Delta t)h_{a}(t_{0}+\Delta t)h_{b}(t_{0})\langle\hat{a}_{a}^{\dagger}\hat{a}_{b}^{\dagger}\hat{a}_{a}\hat{a}_{b}\rangle.
\end{aligned}
\end{equation}
The first and second terms correspond to the second-order self-correlation of the two inputs. The last four correspond to interference, and from these four the first and second are classical (distinguishable), and the third and fourth are quantum due to the indistinguishability of the inputs~\cite{Qian2016HOMIndependentSource,Duan2020HOMSunPhoton}. \par
The second-order correlation function can be computed by integrating the probability $P(t_{0}, \Delta t)$
\begin{equation}
\begin{aligned}
    G^{(2)}(\Delta t)&=\int dt_{0}P(t_{0},\Delta t)\\
    &=G^{(2)}_{a}(\Delta t)+G^{(2)}_{b}(\Delta t)+G^{(2)}_{int}(\Delta t),
\end{aligned}
\end{equation}
where 
\begin{equation}
\begin{aligned}
    G^{(2)}_{a}(\Delta t)&=\langle\hat{a}_{a}^{\dagger}\hat{a}_{a}^{\dagger}\hat{a}_{a}\hat{a}_{a}\rangle t^{2}r^{2}\int dt_{0}|h_{a}(t_{0})h_{a}(t_{0}+\Delta t)|^{2},\\
    G^{(2)}_{b}(\Delta t)&=\langle\hat{a}_{b}^{\dagger}\hat{a}_{b}^{\dagger}\hat{a}_{b}\hat{a}_{b}\rangle \rho^{2}\tau^{2}\int dt_{0}|h_{b}(t_{0})h_{b}(t_{0}+\Delta t)|^{2},\\
    G^{(2)}_{int}(\Delta t)&=\left[\langle\hat{a}_{a}^{\dagger}\hat{a}_{b}^{\dagger}\hat{a}_{b}\hat{a}_{a}\rangle t^{2}\tau^{2}\int dt_{0}|h_{a}(t_{0})h_{b}(t_{0}+\Delta t)|^{2}\right.\\
    &+\langle\hat{a}_{b}^{\dagger}\hat{a}_{a}^{\dagger}\hat{a}_{a}\hat{a}_{b}\rangle r^{2}\rho^{2}\int dt_{0}|h_{a}(t_{0}+\Delta t)h_{b}(t_{0})|^{2}\\
    &+\langle\hat{a}_{b}^{\dagger}\hat{a}_{a}^{\dagger}\hat{a}_{b}\hat{a}_{a}\rangle e^{-i(\theta_{1}+\theta_{2})}\rho r\tau t \int dt_{0}h_{b}^{*}(t_{0})h_{a}^{*}(t_{0}+\Delta t)h_{b}(t_{0}+\Delta t)h_{a}(t_{0})\\
    &\left.+\langle\hat{a}_{a}^{\dagger}\hat{a}_{b}^{\dagger}\hat{a}_{a}\hat{a}_{b}\rangle e^{i(\theta_{1}+\theta_{2})}t\tau r\rho\int dt_{0}h_{a}^{*}(t_{0})h_{b}^{*}(t_{0}+\Delta t)h_{a}(t_{0}+\Delta t)h_{b}(t_{0})\right].
\end{aligned}
\end{equation}
We can split every term that contributes to the interference into two terms to represent the classical and quantum interference
\begin{equation}
\begin{aligned}
    G^{(2)}_{c}(\Delta t)=&\left[\langle\hat{a}_{a}^{\dagger}\hat{a}_{b}^{\dagger}\hat{a}_{b}\hat{a}_{a}\rangle t^{2}\tau^{2}\int dt_{0}|h_{a}(t_{0})h_{b}(t_{0}+\Delta t)|^{2}\right.\\
    &\left.+\langle\hat{a}_{b}^{\dagger}\hat{a}_{a}^{\dagger}\hat{a}_{a}\hat{a}_{b}\rangle r^{2}\rho^{2}\int dt_{0}|h_{a}(t_{0}+\Delta t)h_{b}(t_{0})|^{2}\right],\\
    G^{(2)}_{q}(\Delta t)=&\left[\langle\hat{a}_{b}^{\dagger}\hat{a}_{a}^{\dagger}\hat{a}_{b}\hat{a}_{a}\rangle e^{-i(\theta_{1}+\theta_{2})}\rho r\tau t \int dt_{0}h_{b}^{*}(t_{0})h_{a}^{*}(t_{0}+\Delta t)h_{b}(t_{0}+\Delta t)h_{a}(t_{0})\right.\\
    &\left.+\langle\hat{a}_{a}^{\dagger}\hat{a}_{b}^{\dagger}\hat{a}_{a}\hat{a}_{b}\rangle e^{i(\theta_{1}+\theta_{2})}t\tau r\rho\int dt_{0}h_{a}^{*}(t_{0})h_{b}^{*}(t_{0}+\Delta t)h_{a}(t_{0}+\Delta t)h_{b}(t_{0})\right].
\end{aligned}
\end{equation}\par
At last, we write the normalized second-order correlation function $g^{(2)}(\Delta t)$ which is the ratio of the coincidence counts between the two distinct cases in the Hong-Ou-Mandel interferometer - with and without quantum interference - and considering the contribution from the second-order self-correlation of each input, 
\begin{equation}
    g^{(2)}(\Delta t)=1+\frac{G^{(2)}_{q}(\Delta t)}{G^{(2)}_{a}(\Delta t)+G^{(2)}_{b}(\Delta t)+G^{(2)}_{c}(\Delta t)}.
\end{equation}

\subsection{Coherent state light as input}
For a beam splitter with two coherent state light inputs $|\alpha\rangle_{a}|\beta\rangle_{b}$, we have 
\begin{equation}
\begin{aligned}
    &\langle\hat{a}_{a}^{\dagger}\hat{a}_{a}^{\dagger}\hat{a}_{a}\hat{a}_{a}\rangle=|\alpha|^{4}, \quad
    \langle\hat{a}_{b}^{\dagger}\hat{a}_{b}^{\dagger}\hat{a}_{b}\hat{a}_{b}\rangle=|\beta|^{4},\\
    &\langle\hat{a}_{a}^{\dagger}\hat{a}_{b}^{\dagger}\hat{a}_{b}\hat{a}_{a}\rangle=\langle\hat{a}_{b}^{\dagger}\hat{a}_{a}^{\dagger}\hat{a}_{a}\hat{a}_{b}\rangle=\langle\hat{a}_{b}^{\dagger}\hat{a}_{a}^{\dagger}\hat{a}_{b}\hat{a}_{a}\rangle=\langle\hat{a}_{a}^{\dagger}\hat{a}_{b}^{\dagger}\hat{a}_{a}\hat{a}_{b}\rangle=|\alpha|^{2}|\beta|^{2}.
\end{aligned}
\end{equation}
Then the normalized second-order correlation $g^{(2)}(\Delta t=0)$ of interference at $\Delta t=0$ is 
\begin{equation}
    g^{(2)}(0)=1+\frac{\left[e^{i(\theta_{1}+\theta_{2})}+e^{-i(\theta_{1}+\theta_{2})}\right]rt\tau\rho\cdot|\alpha|^{2}|\beta|^{2}\cdot H_{ab}(0)}{t^{2}r^{2}\cdot|\alpha|^{4}\cdot H_{aa}(0)+\tau^{2}\rho^{2}\cdot|\beta|^{4}\cdot H_{bb}(0)+(t^{2}\tau^{2}+r^{2}\rho^{2})\cdot|\alpha|^{2}|\beta|^{2}\cdot H_{ab}(0)},
\end{equation}
where $H_{ij}(\Delta t=0)=H_{ij}(0)=\int dt_{0}|h_{i}(t_{0})h_{j}(t_{0})|^{2}$ is the temporal wave packet overlap between the two inputs. It is reasonable to set $H_{ij}(0)=1$ when the input modes have perfect overlap and $H_{ij}(0)=0$ when they do not overlap. Thus, for the ideal case where the input modes have perfect temporal overlap
\begin{equation}
   g^{(2)}(0)=1+\frac{\left[e^{i(\theta_{1}+\theta_{2})}+e^{-i(\theta_{1}+\theta_{2})}\right]rt\tau\rho\cdot|\alpha|^{2}|\beta|^{2}}{t^{2}r^{2}\cdot|\alpha|^{4}+\tau^{2}\rho^{2}\cdot|\beta|^{4}+(t^{2}\tau^{2}+r^{2}\rho^{2})\cdot|\alpha|^{2}|\beta|^{2}}. 
\end{equation}
If the two input ports are in balance, i.e., $|\alpha|^{2}=|\beta|^{2}$, and same as before they have perfect temporal wave packet overlap, then we have  
\begin{equation}
    g^{(2)}(0)=1+\frac{\left[e^{i(\theta_{1}+\theta_{2})}+e^{-i(\theta_{1}+\theta_{2})}\right]rt\tau\rho}{t^{2}r^{2}+\tau^{2}\rho^{2}+t^{2}\tau^{2}+r^{2}\rho^{2}}.
\end{equation}

\subsection{Single photon as input}

For a single particle input in each port $|1\rangle_{a}|1\rangle_{b}$, we have 
\begin{equation}
\begin{aligned}
    &\langle\hat{a}_{a}^{\dagger}\hat{a}_{a}^{\dagger}\hat{a}_{a}\hat{a}_{a}\rangle=\langle\hat{a}_{b}^{\dagger}\hat{a}_{b}^{\dagger}\hat{a}_{b}\hat{a}_{b}\rangle=0,\\
    &\langle\hat{a}_{a}^{\dagger}\hat{a}_{b}^{\dagger}\hat{a}_{b}\hat{a}_{a}\rangle=\langle\hat{a}_{b}^{\dagger}\hat{a}_{a}^{\dagger}\hat{a}_{a}\hat{a}_{b}\rangle
    =\langle\hat{a}_{b}^{\dagger}\hat{a}_{a}^{\dagger}\hat{a}_{b}\hat{a}_{a}\rangle=\langle\hat{a}_{a}^{\dagger}\hat{a}_{b}^{\dagger}\hat{a}_{a}\hat{a}_{b}\rangle=1.
\end{aligned}
\end{equation}
Then the normalized second-order correlation $g^{(2)}(\Delta t=0)$ of interference is 
\begin{equation}
    g^{(2)}(0)=1+\frac{\left[e^{i(\theta_{1}+\theta_{2})}+e^{-i(\theta_{1}+\theta_{2})}\right]rt\tau\rho}{t^{2}\tau^{2}+r^{2}\rho^{2}},
\end{equation}
where we assume perfect temporal wave packet overlap. 

\subsection{Effect of splitting ratio}
\begin{figure*}[h]
\includegraphics[width=0.8\textwidth]{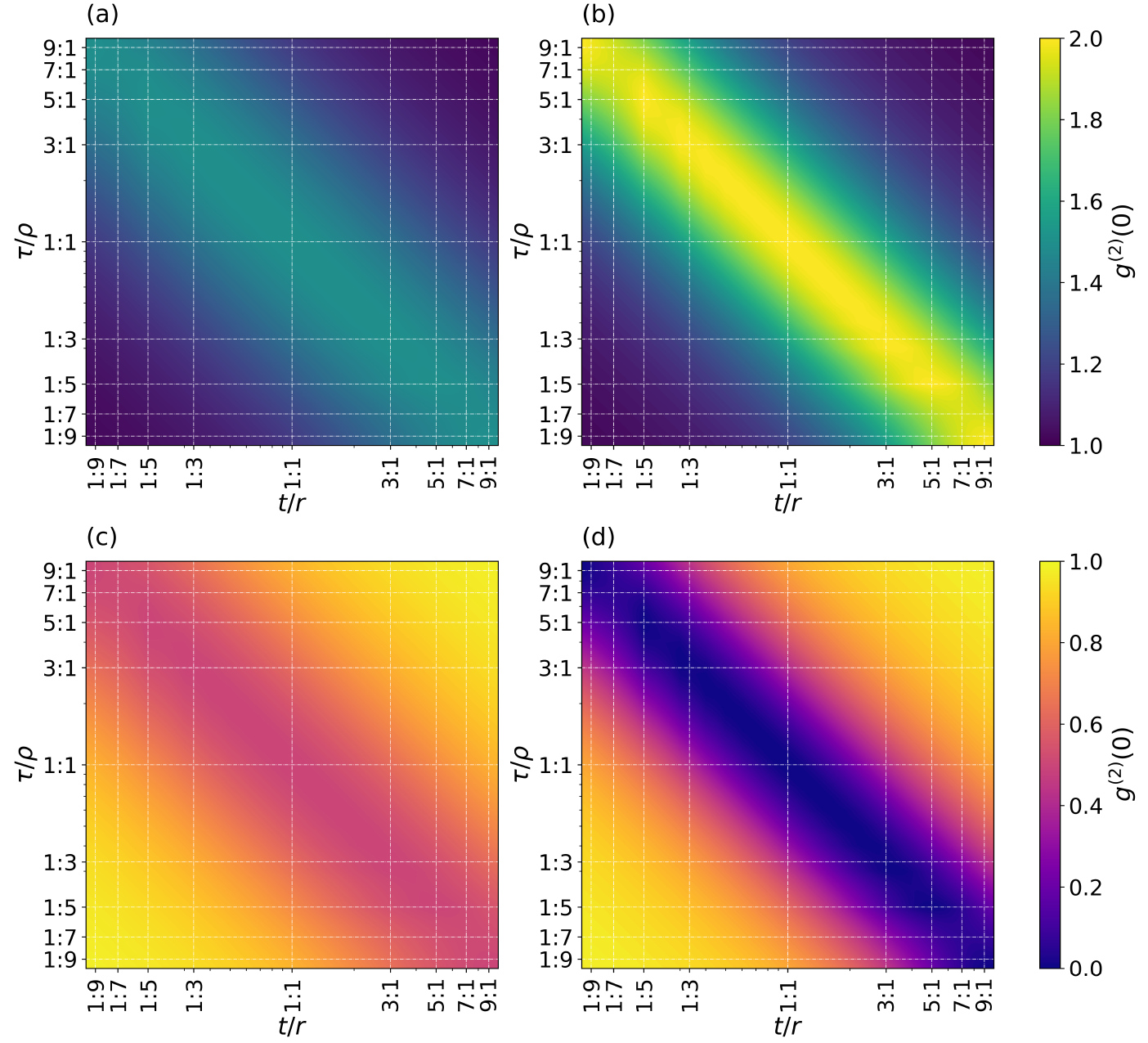} 
\caption{The contour plot of the normalized second-order correlation $g^{(2)}(0)$ for different splitting ratios. (a) and (b) show $g^{(2)}(0)$ for a sum phase $\theta_{+}=0$ with coherent state light input, and single photon input, respectively. (c) and (d) show the same, but for $\theta_{+}=\pi$, where $\theta_{+}=\theta_{1}+\theta_{2}$.}
\label{SMFig4}
\end{figure*}
If we suppose the splitting ratios between $t-r$ and $\tau-\rho$ are $s_{1}$  and $s_{2}$, i.e., $t=s_{1} r$ and $\tau=s_{2} \rho$, we can rewrite the normalized second-order correlation $g^{(2)}(\Delta t=0)$ for coherent state input as
\begin{equation}
    g^{(2)}(0)=1+\frac{\left[e^{i(\theta_{1}+\theta_{2})}+e^{-i(\theta_{1}+\theta_{2})}\right]s_{1}s_{2}}{1+s_{1}^{2}s_{2}^{2}+s_{1}^{2}\frac{r^{2}}{\rho^{2}}+s_{2}^{2}\frac{\rho^{2}}{r^{2}}}.
\end{equation}
Using the formula  
\begin{equation}
    s_{1}^{2}\frac{r^{2}}{\rho^{2}}+s_{2}^{2}\frac{\rho^{2}}{r^{2}}\geq 2s_{1}s_{2}.
\end{equation}
We have  
\begin{equation}
\begin{aligned}
    g^{(2)}(0)\leq 1+\frac{2s_{1}s_{2}}{(1+s_{1}s_{2})^{2}}, \quad\text{for}\ \theta_{1}+\theta_{2}=0,\\
    g^{(2)}(0)\geq 1-\frac{2s_{1}s_{2}}{(1+s_{1}s_{2})^{2}}, \quad\text{for}\ \theta_{1}+\theta_{2}=\pi.
\end{aligned}
\end{equation}
And the normalized second-order correlation $g^{(2)}(\Delta t=0)$ for single particle input is 
\begin{equation}
    g^{(2)}(0)=1+\frac{\left[e^{i(\theta_{1}+\theta_{2})}+e^{-i(\theta_{1}+\theta_{2})}\right]s_{1}s_{2}}{1+s_{1}^{2}s_{2}^{2}}.
\end{equation}
Similarly, 
\begin{equation}
\begin{aligned}
    g^{(2)}(0)= 1+\frac{2s_{1}s_{2}}{1+s_{1}^{2}s_{2}^{2}}, \quad\text{for}\ \theta_{1}+\theta_{2}=0,\\
    g^{(2)}(0)= 1-\frac{2s_{1}s_{2}}{1+s_{1}^{2}s_{2}^{2}}, \quad\text{for}\ \theta_{1}+\theta_{2}=\pi.
\end{aligned}  
\end{equation}\par
The normalized second-order correlations $g^{(2)}(0)$ versus splitting ratio are shown in Fig.~\ref{SMFig4}. We note that even for splitting ratios very far from 1:1, we can still get very high interference contrast, and this relaxes the output matching constraints significantly. \par
If the two input ports are coherent state light and single magnon, the corresponding normalized second-order correlations $g^{(2)}(\Delta t=0)$ are ranging in 
\begin{equation}
\begin{aligned}
    1+\frac{2s_{1}s_{2}}{(1+s_{1}s_{2})^{2}}\leq g^{(2)}(0)\leq 1+\frac{2s_{1}s_{2}}{1+s_{1}^{2}s_{2}^{2}}, \quad\text{for}\ \theta_{1}+\theta_{2}=0,\\
    1-\frac{2s_{1}s_{2}}{(1+s_{1}s_{2})^{2}}\geq g^{(2)}(0)\geq 1-\frac{2s_{1}s_{2}}{1+s_{1}^{2}s_{2}^{2}}, \quad\text{for}\ \theta_{1}+\theta_{2}=\pi.
\end{aligned}  
\end{equation}

\subsection{Unbalanced and mixed input}
For an unbalanced beam splitter with one port coherent state light photon input $|\alpha\rangle_{a}$ and the other one an ideal single magnon input $|1\rangle_{b}$ with perfect temporal wave packet overlap, we have 
\begin{equation}
\begin{aligned}
    &\langle\hat{a}_{a}^{\dagger}\hat{a}_{a}^{\dagger}\hat{a}_{a}\hat{a}_{a}\rangle=|\alpha|^{4},\quad \langle\hat{a}_{b}^{\dagger}\hat{a}_{b}^{\dagger}\hat{a}_{b}\hat{a}_{b}\rangle=0,\\
    &\langle\hat{a}_{a}^{\dagger}\hat{a}_{b}^{\dagger}\hat{a}_{b}\hat{a}_{a}\rangle=\langle\hat{a}_{b}^{\dagger}\hat{a}_{a}^{\dagger}\hat{a}_{a}\hat{a}_{b}\rangle
    =\langle\hat{a}_{b}^{\dagger}\hat{a}_{a}^{\dagger}\hat{a}_{b}\hat{a}_{a}\rangle=\langle\hat{a}_{a}^{\dagger}\hat{a}_{b}^{\dagger}\hat{a}_{a}\hat{a}_{b}\rangle=|\alpha|^{2},
\end{aligned}
\end{equation}
and then the normalized second-order correlation $g^{(2)}(\Delta t=0)$ of this case is 
\begin{equation}
    g^{(2)}(0)=1+\frac{\left[e^{i(\theta_{1}+\theta_{2})}+e^{-i(\theta_{1}+\theta_{2})}\right]rt\tau\rho}{t^{2}r^{2}\cdot|\alpha|^{2}+t^{2}\tau^{2}+r^{2}\rho^{2}}.
\end{equation}\par
\begin{figure*}[h]
\includegraphics[width=0.5\textwidth]{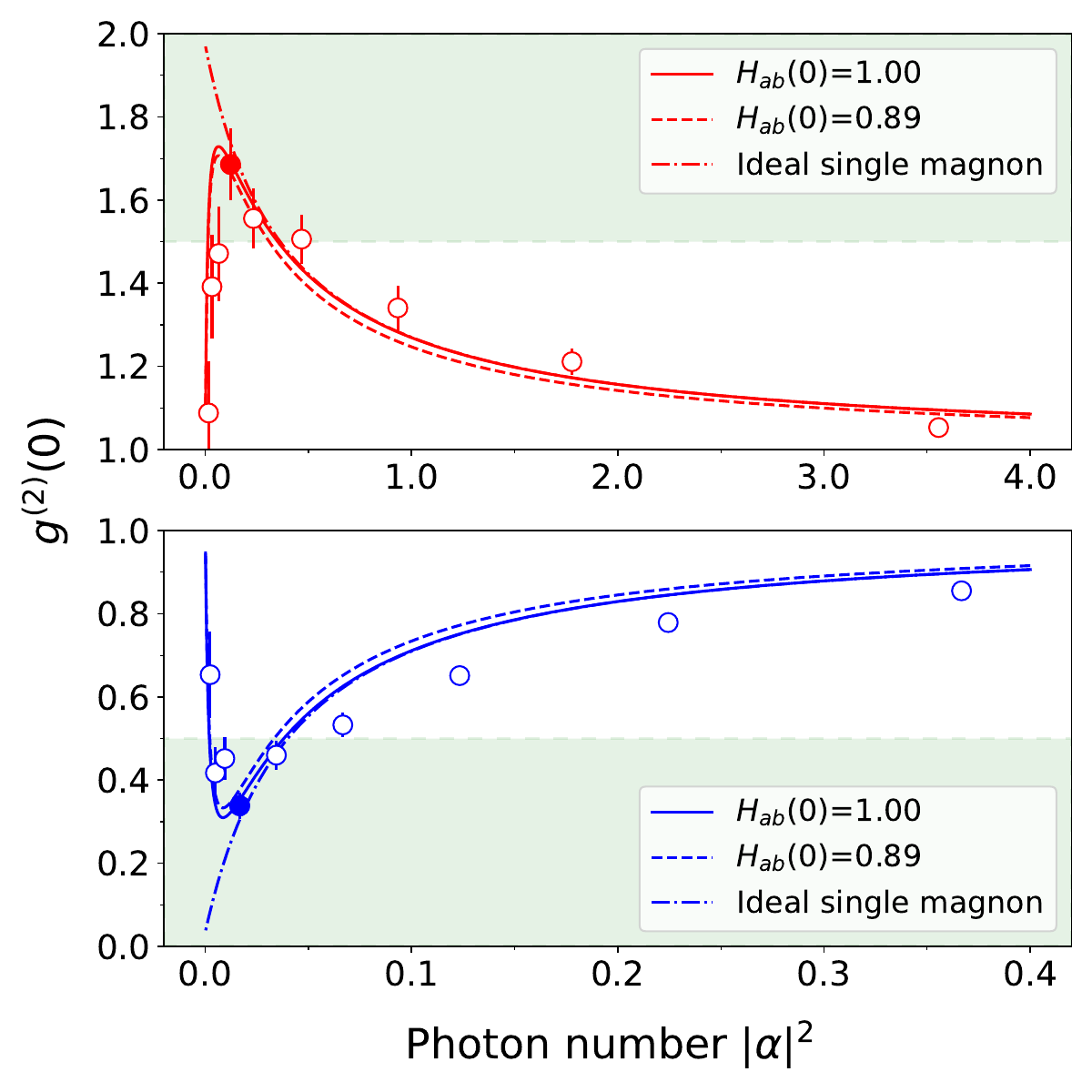} 
\caption{The normalized second-order correlation $g^{(2)}(0)$ versus average coherent state light photon number $|\alpha|^{2}$ for port $a$, and port $b$ with ideal (dot dashed line) and nonideal (solid line) single magnon input. To show the effect of partial wave packet overlap, we also plot the case of nonideal single magnon input with the lowest overlap $H_{ab}(0)=0.89$ (dashed line) appearing in our experiments.}
\label{SMFig5}
\end{figure*}\par
As the single magnon source is not a perfect one, there will be a contribution from multi-magnon events, i.e., $|\psi\rangle_{b}=\sqrt{p_{0}}|0\rangle+\sqrt{p_{1}}|1\rangle+\sqrt{p_{2}}|2\rangle$, thus
\begin{equation}
\begin{aligned}
    &\langle\hat{a}_{a}^{\dagger}\hat{a}_{a}^{\dagger}\hat{a}_{a}\hat{a}_{a}\rangle=|\alpha|^{4},\quad \langle\hat{a}_{b}^{\dagger}\hat{a}_{b}^{\dagger}\hat{a}_{b}\hat{a}_{b}\rangle=2p_{2},\\
    &\langle\hat{a}_{a}^{\dagger}\hat{a}_{b}^{\dagger}\hat{a}_{b}\hat{a}_{a}\rangle=\langle\hat{a}_{b}^{\dagger}\hat{a}_{a}^{\dagger}\hat{a}_{a}\hat{a}_{b}\rangle
    =\langle\hat{a}_{b}^{\dagger}\hat{a}_{a}^{\dagger}\hat{a}_{b}\hat{a}_{a}\rangle=\langle\hat{a}_{a}^{\dagger}\hat{a}_{b}^{\dagger}\hat{a}_{a}\hat{a}_{b}\rangle=|\alpha|^{2}(p_{1}+2p_{2})\approx|\alpha|^{2} p_{1}.
\end{aligned}
\end{equation}
Also, considering the effect of temporal wave packet overlap, the normalized second-order correlation $g^{(2)}(\Delta t=0)$ of this case is 
\begin{equation}
\begin{aligned}
    g^{(2)}(0)&=1+\frac{\left[e^{i(\theta_{1}+\theta_{2})}+e^{-i(\theta_{1}+\theta_{2})}\right]rt\tau\rho\cdot|\alpha|^{2}p_{1}\cdot H_{ab}(0)}{t^{2}r^{2}\cdot|\alpha|^{4}\cdot H_{aa}(0)+\tau^{2}\rho^{2}\cdot 2p_{2}\cdot H_{bb}(0)+(t^{2}\tau^{2}+r^{2}\rho^{2})\cdot|\alpha|^{2}p_{1}\cdot H_{ab}(0)}\\
    &=1+\frac{\left[e^{i(\theta_{1}+\theta_{2})}+e^{-i(\theta_{1}+\theta_{2})}\right]rt\tau\rho\cdot H_{ab}(0)}{t^{2}r^{2}\cdot|\alpha|^{2}/p_{1}+\tau^{2}\rho^{2}g_{b}^{(2)}(0)\cdot p_{1}/|\alpha|^{2}+(t^{2}\tau^{2}+r^{2}\rho^{2})\cdot H_{ab}(0)},
\end{aligned}
\end{equation}
where $g_{b}^{(2)}(0)=2p_{2}/(p_{1}+2p_{2})^{2}\approx 2p_{2}/p_{1}^{2}$ is the normalized second-order self-correlation of the single magnon in port b, and self-overlap of each input temporal wave packet is $H_{aa}(0)=H_{bb}(0)=1$. \par
We define an interference visibility as
\begin{equation}
    V=\left|1-g^{(2)}(0)\right|,
\end{equation}
which we can get from Fig.~\ref{SMFig5}. For ideal single magnon input, we can see the interference visibility rising close to 1 as the average coherent state light photon number drops to levels much lower than one, as Ref.~\cite{Rarity2005Nonclassical} predicts, while for nonideal single magnon input, it rises first and then drops towards zero, as we measured in Fig. 3 of the main text. \par

\section{Temporal wave packet overlap estimation}
The temporal modes of the stored magnons and the incident photons are shaped to be the same as the output mode by the medium and the control laser beam. Thus, we can use the observed output modes of the magnon-photon beam splitter to infer the input modes and their temporal wave packet overlap. We calculate the temporal wave packet overlap $H_{ij}(0)$ for each output port, and all of them are listed in the Table~\ref{Tab1} by using the experimental data shown in Fig.~\ref{SMFig2}. The lowest observed value of $H_{ij}(0)$ is 0.89, and its corresponding normalized second-order correlation is plotted for comparison in Fig.~\ref{SMFig5} (dashed lines). As we can see the difference from the $g^{(2)}(0)$ curve for $H_{ab}(0)=1$ (solid lines) is negligible.
\begin{table}[h]
\centering
\caption{\label{Tab1}Temporal waveform overlap between the output ports.}
\begin{tabularx}{0.8\textwidth}{
    >{\centering\arraybackslash}X
    >{\centering\arraybackslash}X
    >{\centering\arraybackslash}X
    >{\centering\arraybackslash}X
    >{\centering\arraybackslash}X
    >{\centering\arraybackslash}X
    >{\centering\arraybackslash}X
    >{\centering\arraybackslash}X
    >{\centering\arraybackslash}X
    >{\centering\arraybackslash}X}
    \hline\hline
    \multirow{2}{*}{$\theta_{+}$} & \multicolumn{9}{c}{$H_{ij}(0)$}\\
    \cline{2-10}
     & (1)--(2) & (3)--(4) & (5)--(6) & (1)--(3) & (1)--(5) & (3)--(5) & (2)--(4) & (2)--(6) & (4)--(6)\\
    \hline
    0     & 0.97 & 0.98 & 0.92 & 0.98 & 0.96 & 0.94 & 0.98 & 0.98 & 0.97\\
    $\pi$ & 0.97 & 0.97 & 0.89 & 0.99 & 0.99 & 0.98 & 0.95 & 0.90 & 0.89\\
    \hline\hline
\end{tabularx}
\end{table}

\end{document}